\documentclass[twocolumn,preprintnumbers,amsmath,amssymb,prd,
               superscriptaddress]{revtex4}
\usepackage{amsmath,amssymb}
\usepackage{graphicx}% Include figure files
\usepackage{dcolumn}% Align table columns on decimal point
\usepackage{bm}% bold math
\newcommand{\p}{\partial}
\newcommand{\pslash}{p\kern-1ex /}
\newcommand{\lslash}{l\kern-1ex /}
\newcommand{\kslash}{k\kern-1ex /}
\newcommand{\dslash}{\p\kern-1.2ex /}
\newcommand{\Dslash}{{\cal D}\kern-1.5ex /}
\newcommand{\Aslash}{A\kern-1.2ex /}

\newcommand{\tr}{{\rm tr}}

\newcommand{\bea}{\begin{eqnarray}}
\newcommand{\eea}{\end{eqnarray}}

\newcommand{\BAN}{\begin{eqnarray*}}
\newcommand{\EAN}{\end{eqnarray*}}
\newcommand{\Id}{\mbox{1\hspace{-1.2mm}I}}
\begin{document}

\newcommand{\NTU}{
  Physics Department,
  National Taiwan University, Taipei~10617, Taiwan
}

\newcommand{\CQSE}{
  Center for Quantum Science and Engineering,
  National Taiwan University, Taipei~10617, Taiwan
}

\newcommand{\CTS}{
  Center for Theoretical Sciences,
  National Taiwan University, Taipei~10617, Taiwan
}

\preprint{NTUTH-03-505A}

\title{Aspects of Domain-Wall Fermion on the Lattice}

\author{Ting-Wai~Chiu}
\affiliation{\NTU}
\affiliation{\CQSE}
\affiliation{\CTS}

%\collaboration{TWQCD Collaboration}
%\noaffiliation


\begin{abstract}

Two transparent layers are introduced at the boundaries of the fifth dimension, 
for the optimal domain-wall fermions. 
For the quark fields defined in terms of these two transparent layers, 
they obey the usual chiral projection rule in the continuum,
independent of the gauge fields.
Consequently, any observable constructed with the quark fields manifests the
symmetries exactly as those of its counterpart in the continuum.

\end{abstract}

\maketitle

The basic idea of domain-wall fermion (DWF) \cite{Rubakov:bb,Callan:sa}
is to use an infinite set of coupled Dirac fermion fields
\{$ \psi_s (x), s \in (-\infty, \infty) $\} with masses
behaving like a step function $ m(s) = m \theta(s) $ such that the Weyl
fermion state can arise as zero modes bound to the mass defect
at $ s = 0 $. However, if one uses a compact set of masses, then
the boundary conditions of the mass (step) function must
lead to the occurrence of both left-handed and right-handed
chiral fermion fields, i.e., a vector-like theory.
For lattice QCD with DWF \cite{Kaplan:1992bt}, in practice,
one can only use a finite number $ N_s $ of
lattice Dirac fermion fields to set up the domain wall, thus the
chiral symmetry of the quark fields (in the massless limit) is broken.
Therefore, a relevant question is whether one can construct
a domain-wall fermion action such that the effective 4D lattice Dirac
operator has the mathematically optimal chiral symmetry for finite $ N_s $.

In Ref. \cite{Chiu:2002ir},
the optimal domain-wall fermion (ODWF) is constructed
such that the effective 4D lattice Dirac operator
attains the mathematically optimal chiral symmetry for any finite $ N_s $,
exponentially-local for sufficiently smooth gauge backgrounds \cite{Chiu:2002kj},
and independent of the lattice spacing in the fifth dimension.
The basic idea of ODWF is to construct a set of analytical weights,
$ \{ \omega_s, s = 1, \cdots, N_s \} $,
one for each layer in the fifth dimension, such that
the chiral symmetry breaking due to finite $ N_s $
can be reduced to the minimum.
The 4-dimensional effective Dirac operator of massless ODWF is
%\begin{widetext}
\BAN
%\label{eq:odwf_4d}
\begin{aligned}
D &= m_0 [1+ \gamma_5 S_{opt}(H_w) ], \\
S_{opt}(H_w) &= \frac{1-\prod_{s=1}^{N_s} T_s}{1 + \prod_{s=1}^{N_s} T_s}, \quad
T_s = \frac{1-\omega_s H_w}{1+\omega_s H_w},
\end{aligned}
\EAN
%\end{widetext}
which is exactly equal to the Zolotarev optimal rational approximation
of the overlap Dirac operator \cite{Neuberger:1997fp,Narayanan:1995gw}.
That is, $ S_{opt}(H_w) = H_w R_Z(H_w) $, where $ R_Z(H_w)$ is the optimal
rational approximation of $ (H_w^2)^{-1/2} $ \cite{Akhiezer:1992, Chiu:2002eh}.

However, in the original formulation \cite{Chiu:2002ir},
the valence quark propagator cannot be expressed in terms
of the correlation function of the quark fields defined
in terms of the boundary modes, unlike the conventional domain-wall fermion.
In this paper, we solve this problem by introduced two transparent layers
with $ \omega_s = 0 $, as boundary layers appending to the
original action of ODWF such that the quark fields defined in terms of these
two transparent layers obey the usual chiral projection rule in the continuum,
independent of the gauge fields.
Consequently, the valence quark propagator 
can be expressed in terms of the correlation function of the quark fields, 
%$ \langle q(x) \bar q(y) \rangle $, 
and any observable constructed with the quark fields manifests the symmetries 
exactly as those of its counterpart in the continuum. 
The salient feature of a transparent layer (with $ \omega_s = 0 $, and $ T_s = 1 $)
is that its presence does not change the effective 4D Dirac operator.

With two additional transparent layers at $ s=0 $ and $ s=N_s + 1$,
the action of ODWF can be written as 
\bea
\label{eq:ODWF}
\begin{aligned}
{\cal A}_f &= \sum_{s,s'=0}^{N_s+1} \sum_{x,x'}
\bar\psi_{x,s}
\{ (\omega_s D_w + \Id )_{x,x'} \delta_{s,s'} \\
& + (\omega_s D_w - \Id)_{x,x'}
    (P_{-} \delta_{s',s+1} + P_{+} \delta_{s',s-1}) \} \psi_{x',s'} \\
%&\equiv \sum_{s,s'=0}^{N_s+1} \sum_{x,x'}
%        \bar\psi_{x,s} {\cal D}_{x,s;x's'} \psi_{x',s'},
\end{aligned}
\eea
with boundary conditions
\BAN
\label{eq:bc1}
\begin{aligned}
P_{+} \psi_{x,-1} &= - m P_{+} \psi_{x,N_s+1},
\hspace{4mm} m \equiv m_q/(2m_0), \\
\label{eq:bc2}
P_{-} \psi_{x,N_s+2} &= - m P_{-} \psi_{x,0},
\end{aligned}
\EAN
where
$ P_{\pm} = (1 \pm \gamma_5)/2 $,
$ m_q $ is the bare quark mass,
$ D_w $ is the standard Wilson-Dirac operator plus a negative parameter 
$ -m_0 $ ($ 0 < m_0 < 2 $),
%\BAN
%\label{eq:Dw}
%\begin{aligned}
%D_w &= \sum_{\mu=1}^4 \gamma_\mu t_\mu + W - m_0,
%                       \hspace{6mm} m_0 \in (0,2) \\
%(t_\mu)_{x,x'} &= \frac{1}{2} \left[  U_\mu(x) \delta_{x',x+\mu}
%                   - U_\mu^{\dagger}(x') \delta_{x',x-\mu} \right], \\
%W_{x,x'} &= \sum_{\mu=1}^4 \frac{1}{2} \left[ 2 \delta_{x,x'}
%                    - U_\mu(x) \delta_{x',x+\mu}
%                    - U_\mu^{\dagger}(x') \delta_{x',x-\mu} \right],
%\end{aligned}
%\EAN
and the formula for the weights ($ \omega_s, s=1,\cdots,N_s $) is given
in Ref. \cite{Chiu:2002ir}.

Now we define the quark fields in terms of the boundary modes
\bea
\begin{aligned}
\label{eq:q_odwf}
q(x) &= (2m_0)^{-1/2} \left( P_{-} \psi_{x,0} + P_{+} \psi_{x,N_s+1} \right), \\
\label{eq:bar_q_odwf}
\bar q(x) &= (2 m_0)^{-1/2} \left( \bar\psi_{x,0} P_{+} + \bar\psi_{x,N_s+1} P_{-} \right).
\end{aligned}
\eea
In the following, we show that the valence quark propagator in a gauge background is
equal to the correlation function of the quark fields, i.e.,
\bea
\label{eq:Dcmi}
\langle q(x) \bar q(y) \rangle = ( D_c + m_q )^{-1}_{x,y},
\eea
where
\bea
\label{eq:Dc}
D_c &=& 2 m_0 \frac{ 1 + \gamma_5 S_{opt} }{1 - \gamma_5 S_{opt}}, \\
\label{eq:S_opt}
S_{opt} &=& \frac{1-\prod_{s=1}^{N_s} T_s}{1 + \prod_{s=1}^{N_s} T_s}, \\
\label{eq:Ts}
T_s &=& \frac{1-\omega_s H_w}{1+\omega_s H_w}.
\eea
Obviously, the transparent layers (with $ \omega_s = 0 $)
do not change the effective 4D Dirac operator
since $ T_s = 1 $.

The generating functional $ W $ for connected
$n$-point Green's function of the quark fields
is defined as
\bea
\label{eq:ZW}
e^{W[J,\bar J]}
= Z[J,\bar J]
= \frac{ \int e^{ -{\cal A}_g -{\cal A}_f -{\cal A}_{PV}
                  + \bar J q + \bar q J } }
 {\int e^{-{\cal A}_g -{\cal A}_f -{\cal A}_{PV}} },
\eea
where $ \bar J $ and $ J $ are
the Grassman sources of $ q $ and $ \bar q $ respectively, 
$ \int \equiv \int [dU][d\psi][d\bar\psi][d\phi][d\bar\phi] $,
$ {\cal A}_g $ is the gauge action,
$ {\cal A}_{PV} $ is the action of the Pauli-Villars fields 
$ \{ \bar \phi_s, \phi_s \} $ with $ m_q = 2 m_0 $, i.e., 
\BAN
\begin{aligned}
{\cal A}_{PV} &= \sum_{s,s'=0}^{N_s+1} \sum_{x,x'}
\bar\phi_{x,s}
\{ (\omega_s D_w + \Id )_{x,x'} \delta_{s,s'} \\
& + (\omega_s D_w - \Id)_{x,x'}
    (P_{-} \delta_{s',s+1} + P_{+} \delta_{s',s-1}) \} \phi_{x',s'} \\
%&\equiv \sum_{s,s'=0}^{N_s+1} \sum_{x,x'}
%        \bar\psi_{x,s} {\cal D}_{x,s;x's'} \psi_{x',s'},
\end{aligned}
\EAN
with boundary conditions
\BAN
\begin{aligned}
P_{+} \phi(x,-1) &= - P_{+} \phi(x,N_s+1), \\
P_{-} \phi(x,N_s+2) &= - P_{-} \phi(x,0).
%\label{eq:bc}
\end{aligned}
\EAN

First we evaluate the fermionic integrals in (\ref{eq:ZW}).
Using $ \gamma_5 P_{\pm} = \pm P_{\pm} $,
$ P_{+} + P_{-} = 1 $, and $ H_w = \gamma_5 D_w $,
we can rewrite (\ref{eq:ODWF}) as
\bea
\label{eq:A_3a}
\begin{aligned}
{\cal A}_f &=
  (m + 1) \bar\psi_0 \gamma_5 P_{+} \psi_{N_s+1} \\
- & \bar\psi_0\gamma_5(P_{-}\psi_0 + P_{+} \psi_{N_s+1})
   +\bar\psi_0\gamma_5(P_{-}\psi_1 + P_{+} \psi_0) \\
& + \sum_{s=1}^{N_s} \{ \bar\psi_s\gamma_5(\omega_s H_w - \Id)(P_{-}\psi_s + P_{+} \psi_{s-1}) \\
& \hspace{10mm}
  + \bar\psi_s\gamma_5(\omega_s H_w + \Id)(P_{-}\psi_{s+1} + P_{+} \psi_s) \}  \\
& - \bar\psi_{N_s+1} \gamma_5 (P_{-} \bar\psi_{N_s+1} + P_{+} \psi_{N_s} )  \\
& + \bar\psi_{N_s+1} \gamma_5 (P_{-}\psi_0 + P_{+} \psi_{N_s+1} ) \\
& -(m + 1)\bar\psi_{N_s+1} \gamma_5 P_{-} \psi_0,
\end{aligned}
\eea
where all indices are suppressed except the index in the
5-th dimension.
Next we define
\BAN
%\label{eq:o1}
\begin{aligned}
& \eta_0 = P_{-} \psi_0 + P_{+} \psi_{N_s+1}, \
\bar\eta_0 = - \bar\psi_0 \gamma_5, \\
%\label{eq:o2}
& \eta_s = P_{-} \psi_s + P_{+} \psi_{s-1}, \
\bar\eta_s = \bar\psi_s \gamma_5 (\omega_s H_w - 1),  \\
%\label{eq:o3}
& \eta_{N_s+1} = P_{-} \psi_{N_s+1} + P_{+} \psi_{N_s}, \
\bar\eta_{N_s+1} = -\bar\psi_{N_s+1} \gamma_5,
\end{aligned}
\EAN
where the index $ s $ in the second line runs from 1 to $ N_s $,
and the inverse transform
\BAN
%\label{eq:o1i}
\begin{aligned}
&
\psi_0 = P_{-} \eta_0 + P_{+} \eta_1, \
\bar\psi_0 = - \bar\eta_0 \gamma_5, \\
%\label{eq:o2i}
&
\psi_s = P_{-} \eta_s + P_{+} \eta_{s+1}, \
\bar\psi_s = \bar\eta_s (\omega_s H_w - 1)^{-1} \gamma_5, \\
%\label{eq:o3i}
&
\psi_{N_s+1} = P_{-} \eta_{N_s+1} + P_{+} \eta_0, \
\bar\psi_{N_s+1} = -\bar\eta_{N_s+1} \gamma_5.
\end{aligned}
\EAN
Then the action (\ref{eq:A_3a}) can be rewritten as
\bea
\begin{aligned}
\label{eq:A_3b}
{\cal A}_f
&= \bar\eta_0 ( P_{-} - m P_{+} ) \eta_0
  - \bar\eta_0 \eta_1 \\
&   + \sum_{s=1}^{N_s} \left\{
    \bar\eta_s \eta_s - \bar\eta_s T_s^{-1} \eta_{s+1}
              \right\}              \\
&   + \bar\eta_{N_s+1} \eta_{N_s+1}
    - \bar\eta_{N_s+1} ( P_{+} - m P_{-} ) \eta_0.
\end{aligned}
\eea
Thus the fermionic integral in the numerator of (\ref{eq:ZW})
can be written as
\bea
\begin{aligned}
& \hspace{4mm} \int [d\bar\psi] [d\psi]
     e^{-{\cal A}_f[\psi,\bar\psi] + \bar J q + \bar q J } \\
& = {\cal J}\int [d\bar\eta] [d\eta]
     e^{-{\cal A}_f[\eta,\bar\eta] + \bar J' \eta_0 - \bar\eta_0 P_{+} J'
                 + \bar\eta_{N_s+1} P_{-} J' },
\label{eq:Odet}
\end{aligned}
\eea
where $ J'=(2m_0)^{-1/2} J $ and $ \bar J'=(2 m_0)^{-1/2} \bar J $,
%\bea
%\begin{aligned}
%& \hspace{4mm} \bar J q + \bar q J \\
%& =  \bar J' ( P_{-} \psi_0 + P_{+} \psi_{N_s+1} )
%   + ( \bar\psi_0 P_{+} + \bar\psi_{N_s+1} P_{-} ) J' \\
%& =  \bar J' \eta_0 - \bar\eta_0 P_{+} J + \bar\eta_{N_s+1} P_{-} J'
%\label{eq:jq}
%\end{aligned}
%\eea
and ${\cal J}$ is the Jacobian of the transformation,
\BAN
\label{eq:j}
{\cal J}
%=\prod_{s=1}^{N_s} \det \left[\gamma_5 (\omega_s H_w-1) \right].
=\prod_{s=1}^{N_s} \det (\omega_s H_w-1).
\EAN
Now using the Grassman integral formula
\BAN
\label{eq:G_integral}
\int d \bar\chi d \chi \ e^{-\bar\chi M \chi + \bar v \chi + \bar \chi v}
= e^{\bar v M^{-1} v} \det M,
\EAN
one can easily evaluate the Grassman integrals in (\ref{eq:Odet}),
by integrating $ (\eta_{s}, \bar\eta_{s}) $ successively from
$ s = N_s + 1 $ to $ s = 0 $.
Explicitly, after integrating $ (\eta_{N_s+1}, \bar\eta_{N_s+1}) $,
(\ref{eq:Odet}) becomes
\BAN
\begin{aligned}
& {\cal J}\int \prod_{s=0}^{N_s}[d \bar\eta_s] [d \eta_s]
            \exp \{ - \bar\eta_0 ( P_{-} - m P_{+} ) \eta_0
                    + \bar\eta_0 \eta_1 \\
&  \hspace{20mm}    + \bar J' \eta_0 - \bar\eta_0 P_{+} J' \\
&  \hspace{10mm}
-\bar\eta_{N_s} \eta_{N_s}   \\
&  \hspace{10mm}  +\bar\eta_{N_s} T_{N_s}^{-1}
                 [ (P_{+} - m P_{-}) \eta_0 + P_{-} J' ] \},
\end{aligned}
\EAN
then integrating $(\eta_{N_s},\bar\eta_{N_s})$, it becomes
\BAN
\begin{aligned}
& {\cal J}\int \prod_{s=0}^{N_s-1}[d \bar\eta_s] [d \eta_s]
   \exp \{ - \bar\eta_0 ( P_{-} - m P_{+} ) \eta_0
                          + \bar\eta_0 \eta_1   \\
& \hspace{20mm} + \bar J' \eta_0 - \bar\eta_0 P_{+} J'  \\
& \hspace{10mm}
-\sum_{s=1}^{N_s-2} (\bar\eta_s \eta_s - \bar\eta_s T_s^{-1} \eta_{s+1} )
-\bar\eta_{N_s-1} \eta_{N_s-1} \\
& \hspace{10mm}
   +\bar\eta_{N_s-1} T_{N_s-1}^{-1} T_{N_s}^{-1}
    [(P_{+} - m P_{-}) \eta_0 + P_{-} J' ] \}.
\end{aligned}
\EAN
Subsequent integrations over $ (\eta_{N_s-2}, \bar\eta_{N_s-2}) $
up to $ (\eta_1, \bar\eta_1) $ are similar to the above integration,
and the result is
\BAN
\begin{aligned}
& {\cal J}\int [d \bar\eta_0] [d \eta_0]
    \exp \{ - \bar\eta_0 ( P_{-} - m P_{+} ) \eta_0
            + \bar J' \eta_0 - \bar\eta_0 P_{+} J' \\
& \hspace{10mm}
   +\bar\eta_0 \prod_{s=1}^{N_s} T_s^{-1}
   [ (P_{+} - m P_{-}) \eta_0 + P_{-} J' ] \}
\end{aligned}
\EAN
Finally, integrating $ (\eta_0,\bar\eta_0) $ gives
\BAN
\label{eq:int_f}
\begin{aligned}
&
{\cal J} \det \left[ (P_{-}- m P_{+})
         - T^{-1} (P_{+}- m P_{-}) \right] \times
\\
&
e^{ \bar J' \left[ \left(-P_{+}+ T^{-1} P_{-} \right)^{-1}
    \left(P_{-}-T^{-1} P_{+} \right) + m \right]^{-1} J'
  }
\end{aligned}
\EAN
where $ T^{-1} = \prod_{s=1}^{N_s} T_s^{-1} $.
Using the simple identity
\BAN
%\begin{aligned}
  \left(-P_{+}+ T^{-1} P_{-} \right)^{-1} \left( P_{-}- T^{-1} P_{+} \right)
= \frac{1 + \gamma_5 S_{opt}}{1 - \gamma_5 S_{opt}},
%= (2 m_0)^{-1} D_c,
%\end{aligned}
\EAN
the above result becomes
\bea
\label{eq:ZW_N_f}
{\cal K} \det [ (D_c + m_q)(2 m_0)^{-1} ]
  \exp \{ \bar J  (D_c + m_q )^{-1} J \},
\eea
where $ D_c $ is defined in (\ref{eq:Dc}), and
\BAN
\label{eq:K}
{\cal K} = \prod_{s=1}^{N_s} \det (\omega_s H_w-1) \cdot 
           \det \left[ (-P_{+} + T^{-1} P_{-}) \right].
\EAN
Setting $ \bar J = J = 0 $ in (\ref{eq:ZW_N_f}), we obtain
the result for the fermionic integral in the denominator of (\ref{eq:ZW}),
\bea
\label{eq:ZW_D_f}
\int [d\bar\psi] [d\psi] e^{-{\cal A}_f} =
{\cal K} \det [ (D_c + m_q)(2 m_0)^{-1} ].
\eea

Next, we evaluate the integrals over the Pauli-Villars fields in
(\ref{eq:ZW}). Using the Gaussian integration formula for the
boson fields, and following the procedures similar to
above for the fermion fields, we obtain
\bea
\label{eq:int_PV}
\int [d \bar\phi][d \phi] e^{-{\cal A}_{PV}}
= \pi^{N_s+2} {\cal K}^{-1}  \det [ 1 + D_c/(2m_0) ]^{-1}.
\eea

Substituting (\ref{eq:ZW_N_f})-(\ref{eq:int_PV})
into (\ref{eq:ZW}), we have
\bea
\label{eq:ZW_odwf}
e^{W[J,\bar J]} =
\frac{ \int [dU] e^{-{\cal A}_g} \det D(m_q) e^{ \bar J ( D_c + m_q )^{-1} J } }
     { \int [dU] e^{-{\cal A}_g} \det D(m_q) },
\eea
where
\bea
\begin{aligned}
D(m_q) &= (D_c + m_q) [ 1 + D_c/(2m_0) ]^{-1} \\
&= m_q + (m_0 - m_q/2) [1+ \gamma_5 S_{opt}(H_w) ],
\end{aligned}
\eea
is the effective 4D operator of ODWF, in which
$ S_{opt}(H_w) $ is exactly equal to the Zolotarev optimal
rational approximation of the sign function in the 
overlap Dirac operator \cite{Neuberger:1997fp}. 
That is, $ S_{opt}(H_w) = H_w R_Z(H_w) $, where $ R_Z(H_w)$ is the optimal
rational approximation of $ (H_w^2)^{-1/2} $ \cite{Akhiezer:1992, Chiu:2002eh}.
In the limit $ N_s \to \infty $, $ S_{opt}(H_w) = H_w (H_w^2)^{-1/2} $, 
$ \gamma_5 S_{opt}(H_w) \equiv V $ satisfying $ V^\dagger = \gamma_5 V \gamma_5 = V^{-1} $. 
Then $ D_c $ (\ref{eq:Dc}) becomes $ D_c = 2 m_0 (1+V)(1-V)^{-1} $ which is chirally symmetric, 
and $ D(0) = D_c ( 1 + D_c )^{-1} = m_0 (1+V) $ is exactly equal to 
the overlap Dirac operator \cite{Neuberger:1997fp}, 
satisfying the Ginsparg-Wilson relation
\BAN
D(0) \gamma_5 + \gamma_5 D(0) = \frac{1}{m_0} D(0) \gamma_5 D(0).
\EAN

The quark propagator can be obtained by differentiating
$ W[J,\bar J] $ with respect to $ J $ and $ \bar J $,
\bea
\begin{aligned}
& \hspace{4mm} \langle q(x) \bar q(y) \rangle =
 \left.
-\frac{\delta^2 W[J,\bar J]}{\delta \bar J(x) \delta J(y)}
\right|_{J=\bar J=0} \\
&= \frac{ \int [dU] e^{-{\cal A}_g} \det [D(m_q)]
           (D_c + m_q)^{-1}_{x,y} }
         { \int [dU] e^{-{\cal A}_g} \det [D(m_q)] },
\end{aligned}
\eea
which reduces to (\ref{eq:Dcmi}) for a background gauge field.

In general, any observable involving quark fields can be obtained
from $ Z[J, \bar J] $ by differentiation, and it possesses
the symmetries exactly the same as its counterpart in the continuum.
For example, the current-current correlator
\BAN
\begin{aligned}
& \langle \bar d(x) \gamma_4 P_{-} s(x)
            \bar s(0) \gamma_4 P_{-} d(0) \rangle  \\
%            \hspace{12mm}
%     P_{\pm} = \frac{1}{2}( 1 \pm \gamma_5 )   \\
=&
    \left. \frac{\delta}{\delta J_d(x)} \gamma_4 P_{-}
             \frac{\delta}{\delta \bar J_s(x)}
             \frac{\delta}{\delta J_s(0)} \gamma_4 P_{-}
             \frac{\delta}{\delta \bar J_d(0)}
             Z[J, \bar J ] \right|_{0}  \\
=& \frac{\int [dU] e^{-{\cal A}_G } \prod_f \det D(m_f) O_K(x)}
         {\int [dU] e^{-{\cal A}_G} \prod_f \det D(m_f)}
\end{aligned}
\EAN
where
\BAN
O_K(x)
= -\tr [(D_c + m_d)^{-1}_{0,x} \gamma_4 P_{-}
       (D_c + m_s)^{-1}_{x,0} \gamma_4 P_{-} ].
\EAN
Obviously, the $ V-A $ structure of the left-handed quark currents
is preserved exactly. This is one of the basic motivations to
introduce the transparent layers with $ \omega_s = 0 $.
On the other hand, if one uses the Ginsparg-Wilson Dirac operator
(satisfying $ D \gamma_5 + \gamma_5 D = D \gamma_5 D $)
to construct the quark action,
\BAN
\begin{aligned}
{\cal A}_F &= \sum_{x,y} \bar q(x) D_{x,y} q(y)    \\
&= \sum_{x,y} [ \bar q(x) P_{+} ( D \hat P_{-})_{x,y} q(y) +
             \bar q(x) P_{-} ( D \hat P_{+})_{x,y} q(y) ]
\end{aligned}
\EAN
where $ P_{\pm} = (1 \pm \gamma_5)/2 $ and
$ \hat{P}_{\pm} = \frac{1}{2}[ 1 \pm \gamma_5 (1 - D) ] $
are the chiral projectors for $ \bar q $ and $ q $ respectively.
Then $\bar d(x) \gamma_\mu \hat{P}_{-} s(x) \ne \bar d(x) P_{+} \gamma_\mu s(x)$,
and the left-handed quark current does not manifest the $ V-A $ structure.
Consequently, these left-handed quark currents explicitly breaks the $ SU_L(2) $ 
gauge symmetry by $ O(a) $ effect \cite{Fujikawa:2002vj},  
and the discrete CP symmetry in chiral gauge theories with Ginsparg-Wilson fermion 
is explicitly broken by $ O(a) $ effect \cite{Hasenfratz:2001bz}.

Even though the transparent layers (with $ \omega_s = 0 $)
are introduced as the boundary layers for defining the quark fields
such that any observable involving the quark fields
manifest the symmetries exactly as those of
its counterpart in the continuum,
in practice, one does not need to keep these transparent layers
in the dynamical simulations of QCD, since their presence
does not change the fermion determinant at all.
Moreover, the valence quark propagator also can be obtained
without using the transparent layers.
In the original ODWF action \cite{Chiu:2002ir},
the quark fields are defined by the boundary modes at $ s = 1$
and $ s = N_s $ similar to (\ref{eq:q_odwf}).
Then one can obtain the valence quark propagator
by solving the following linear system
\bea
\label{eq:DY}
{\cal D}(m_q) |Y \rangle = {\cal D}(2m_0) B^{-1} |\mbox{source vector} \rangle
\eea
%where $B $ and $ B^{-1} $ are defined as
%\BAN
%\label{eq:B}
%B_{x,s;x',s'} &=& \delta_{x,x'} (P_{-}\delta_{s,s'}+P_{+}\delta_{s-1,s'}), \\
%\label{eq:BI}
%B^{-1}_{x,s;x',s'} &=& \delta_{x,x'}(P_{-}\delta_{s,s'}+P_{+}\delta_{s+1,s'}),
%\EAN
where $ B^{-1}_{x,s;x',s'} = \delta_{x,x'}(P_{-}\delta_{s,s'}+P_{+}\delta_{s+1,s'}) $
with periodic boundary conditions in the fifth dimension.
Then the solution of (\ref{eq:DY}) gives the valence quark propagator
\BAN
\label{eq:v_quark}
(D_c + m_q)^{-1}_{x,x'} = \left( 2 m_0 - m_q \right)^{-1} \left[ (BY)_{x,1;x',1} - \delta_{x,x'} \right].
\EAN

Nevertheless, the transparent layers with $ \omega_s = 0 $
turn out to play a crucial role in the derivations of some analytical
results which would be difficult to obtain otherwise.
An example is the derivation of the axial Ward identity for lattice QCD 
with ODWF \cite{Chen:2012jy}.
Obviously, one can insert any numbers of transparent layers
at any locations along the 5-th dimension.
This opens new possibilities to tackle problems
in domain-wall fermions.

%
%\begin{acknowledgments}
  This work is supported in part by the National Science Council
  (Grant No. NSC99-2112-M-002-012-MY3), and NTU-CQSE (Grant No. 10R80914-4).

%\end{acknowledgments}

\end{document}